\documentclass[pre,twocolumn,aps]{revtex4-1}
\usepackage{epsfig}

\begin{document}
\title{Beyond power-law density scaling: Theory, Simulation, and Experiment}
\author{Lasse B{\o}hling$^*$, Trond S. Ingebrigtsen$^*$, A. Grzybowski$^o$, M. Paluch$^o$, Jeppe C. Dyre$^*$, and Thomas B. Schr{\o}der$^*$}
\affiliation{$^*$DNRF Centre ``Glass and Time'', IMFUFA, Department of Sciences, Roskilde University, Postbox 260, DK-4000 Roskilde, Denmark.}
\affiliation{$^o$Institute of Physics, University of Silesia, ul. Uniwersytecka 4, 40-007 Katowice, Poland.}

\date{\today}

\begin{abstract}
Supercooled liquids are characterized by relaxation times that increase dramatically by cooling or compression. Many liquids have been shown to obey power-law density scaling, according to which the relaxation time is a function of density to some power over temperature. We show that power-law density scaling breaks down for larger density variations than usually studied. This is demonstrated by simulations of the Kob-Andersen binary Lennard-Jones mixture and two molecular models, as well as by experimental results for two van der Waals liquids. A more general form of density scaling is derived, which is consistent with results for all the systems studied. 
An analytical expression for the scaling function for liquids of particles interacting via generalized Lennard-Jones potentials is  derived and shown to agree very well with simulations. This effectively reduces the problem of understanding the viscous slowing down from being a quest for a function of two variables to a search for a single-variable function.
\end{abstract}
\maketitle

The relaxation time of a supercooled liquid increases markedly upon cooling, in some cases by a factor of ten or more when temperature decreases by just one percent.\cite{kau48,har76,bra85,gut95,edi96,ang00,alb01,deb01,bin05,sci05,dyr06}  This phenomenon lies behind glass formation, which takes place when a liquid upon cooling is no longer able to equilibrate on laboratory time scales due to its extremely long relaxation time. It has long been known that increasing pressure at constant temperature increases the relaxation time in much the same way as does cooling at ambient pressure. Only during the last decade, however, have large amounts of data become available on how the relaxation time varies with temperature and density. Following pioneering works by T{\"o}lle,\cite{tol01} it was demonstrated by Dreyfus {\it et al.},\cite{dre03} Alba-Simionesco {\it et al.},\cite{alb}, as well as Casalini and Roland,\cite{cas04} that for many liquids and polymers the relaxation time  is a function of a single variable. 
Roland {\it et al.}\cite{rol05} reviewed the field, and demonstrated that for a large number of molecular liquids and polymers the relaxation time is a function of  $\rho^\gamma/T$, where $\gamma$ is an empirical material-dependent parameter. 
For recent works on this ``power-law density scaling'' or ``thermodynamic scaling'' see, e.g., Refs.~\onlinecite{rol10,flo11,lop11,gun11}. The consensus is now that van der Waals liquids and most polymers conform to the scaling, whereas hydrogen-bonding liquids disobey it.

\begin{figure}
\begin{center} 
   \includegraphics[width=0.45\textwidth]{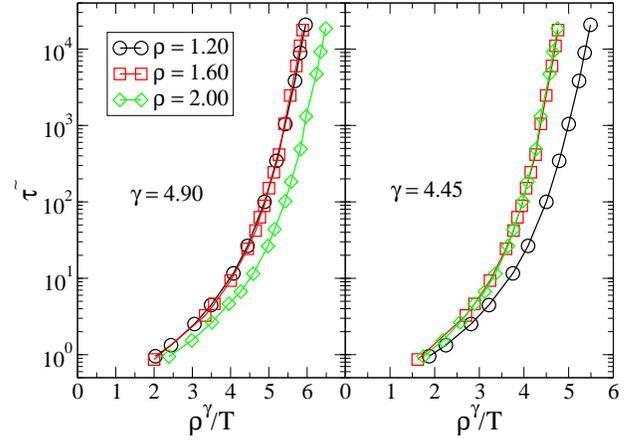}
\end{center}
   \caption{
Breakdown of power-law density scaling for the reduced structural relaxation time $\tilde{\tau}_{\alpha}$ in the KABLJ mixture, $\tilde{\tau}_{\alpha} \equiv \rho^{1/3}(k_BT/m)^{1/2} \tau_{\alpha}$, where $\tau_{\alpha}$ is the time at which the self-intermediate scattering function ($F_s(q,t)$, $q=7.25(\rho/1.2)^{1/3}$) for the A particles has decayed to $1/e$. Molecular dynamics (MD) simulations in the NVT ensemble ($N=1000$) were performed using RUMD, a MD package optimized for state of the art GPU computing.\cite{RUMD} $\tilde{\tau}_{\alpha}$  is plotted for three isochores as a function of the density-scaling variable $\rho^\gamma/T$, where $\gamma$ is an empirical scaling parameter. Left panel: $\gamma=4.90$ collapses the data for the two lowest densities. Right panel:  $\gamma=4.45$ collapses the data for the two highest densities. It is not possible to find a single exponent that collapses all the data;  even though the two exponents differ by only 10\%, power-law scaling with a single exponent clearly fails.}
    \label{fig:ds_break}
\end{figure}

A standard model in simulation studies of viscous liquids is the Kob-Andersen binary Lennard-Jones (KABLJ) mixture,\cite{KA} which can be cooled to a highly viscous state and only crystallizes for extraordinarily long simulations.\cite{tox09} The system consists of 80\% large Lennard-Jones (LJ) particles (A) interacting  strongly with 20\% smaller LJ particles (B).\cite{KABLJ} The KABLJ mixture was shown by Coslovich and Roland\cite{cos09} to obey power-law density scaling to a good approximation  with $\gamma=5.1$  for the density range $\rho\equiv N/V=1.15$ to $\rho=1.35$, whereas Pedersen {\it et al.}\cite{urp10} used the slightly different value $\gamma=5.16$ to scale the density range 1.1 to 1.4. Figure \ref{fig:ds_break} demonstrates, however, that power-law density scaling breaks down when considering a larger density range. Relaxation time data for the isochores  $\rho=1.2$ and $\rho=1.6$ collapse very well using $\gamma = 4.90$, whereas the isochores $\rho=1.6$ and $\rho=2.0$ collapse using $\gamma = 4.45$; in both cases the third isochore deviates significantly. In the following we  show that power-law density scaling is an approximation to a more general form of scaling, which is derived from the theory of isomorphs.\cite{paperIV,paperV}  We further show that given that the two lowest densities of Fig. 1 obey power-law density scaling with $\gamma = 4.90$, the isomorph theory predicts that the two highest densities scales with $\gamma = 4.45$, as seen in  Fig.~\ref{fig:ds_break}.





What causes power-law density scaling and its break-down for large density variations? A justification of density scaling may be given by reference to inverse power-law (IPL) potentials ($\propto r^{-n}$) where $r$ is the distance between particles. For such unrealistic, purely repulsive systems density scaling is rigorously obeyed with $\gamma=n/3$ \cite{ipl}. Assuming that power-law density scaling reflects some sort of underlying effective power-law potential, the scaling exponent $\gamma$ can be found from the NVT equilibrium fluctuations in the potential energy $U$ and the virial $W=pV - NkT$ (IPL potentials have $W=(n/3)U$) as follows:
\begin{equation}
 \gamma  = \frac{\langle \Delta W \Delta U  \rangle}{\langle (\Delta U)^2 \rangle}. \label{eq:fluctgamma}
\end{equation}
This was confirmed for the KABLJ mixture by Refs. \onlinecite{cos09} and \onlinecite{urp10}. Ref. \onlinecite{urp10} further supported this ``hidden scale invariance'' explanation by demonstrating that for the investigated density range the dynamics and structure of the KABLJ mixture is accurately reproduced by an IPL mixture with exponent chosen in accordance with Eq.~(\ref{eq:fluctgamma}).

The hidden scale invariance is not just a feature of the KABLJ mixture but of ``strongly correlating liquids'' in general.\cite{ped08,paperI,paperII,paperIII} These are defined by having strong correlations between equilibrium NVT fluctuations of the potential energy and the virial (correlation coefficients larger than 0.9). Also molecular models can be strongly correlating; examples include the Lewis-Wahnstrom model of ortho-terphenyl and an asymmetric dumbell model. Both models are strongly correlating and obey power-law density scaling with exponents consistent with Eq.~(\ref{eq:fluctgamma}) for density increases of 8\% and 16\% respectively.\cite{sch09} Very recently Gundermann {\it et al.}\cite{gun11} investigated the van der Waals liquid tetramethyl-tetraphenyl-trisiloxane, and gave the first experimental confirmation of the relation between the power-law scaling exponent and Eq.~(\ref{eq:fluctgamma}).

For any potential that is \emph{not} an IPL potential the exponent $\gamma$ as calculated from Eq.~(\ref{eq:fluctgamma}) depends on the state point. Power-law density scaling corresponds to disregarding this state-point dependence. It is thus not surprising that power-law density scaling breaks down when large density changes are involved, but interestingly a more fundamental and robust form of scaling can be derived as we now proceed to show. 



Strongly correlating liquids have ``isomorphs'' in their phase diagram, which are curves along which structure and dynamics in reduced units are invariant to a good approximation.\cite{paperIV,paperV} The invariance of structure implies invariance of the configurational (``excess'') entropy, $S_{ex}$, thus the isomorphs are nothing but the configurational adiabats. Ref. \onlinecite{paperIV} discussed in detail the consequences of a liquid having isomorphs and also showed that a liquid is strongly correlating if and only if it has isomorphs to a good approximation. 
The precise statistical-mechanical definition of an isomorph\cite{paperIV} is that this is an equivalence class of state points, where two state points are termed equivalent (``isomorphic'') if all pairs of physically relevant micro-configurations of the two state points, which  trivially scale into one another, have proportional configurational Boltzmann's factors. From this single assumption several predictions can be derived, including isomorph invariance of structure and dynamics in reduced units and that jumps between isomorphic state points take the system instantaneously to equilibrium.\cite{paperIV}

Letting $\mathbf R$ denote a micro-configuration (all particle coordinates) of a reference state point $(\rho_*, T_*)$, the condition for state point $(\rho, T)$ to be isomorphic with $(\rho_*, T_*)$, i.e., the proportionality of pairs of Boltzmann's factors, can, by taking the logarithm and rearranging, be expressed as:
\begin{equation}
   U\left( \tilde\rho^{-\frac{1}{3}}  \mathbf R \right) = 
   \frac{T}{T_*} U\left(\mathbf R \right) + K, ~~~\tilde\rho\equiv\rho/\rho_* \label{Eq:DirectIsomorph}
\end{equation} 
where $K$ is a constant that only depends on the two state points. Equation~(\ref{Eq:DirectIsomorph}) is the basis of the so called \emph{direct isomorph check}: \cite{paperIV} a) draw micro-configurations $\mathbf R$ from a simulation at $(\rho_*, T_*)$, b) evaluate the potential energies of these configurations scaled to density $\rho$, and plot them in a scatter-plot against the potential energies at $\rho_*$. If a state point $(\rho,T)$  exists that to a good approximation is isomorphic with $(\rho_*,T_*)$, this scatter plot will be close to a straight line and $T$ can be found from the slope.

In the following we consider systems where the interaction potential between particles $i$ and $j$ is given by a sum of inverse power laws: 
\begin{equation}
  \phi_{ij}(r_{ij}) = \sum_{n} \epsilon_{n,ij}\left(\frac{\sigma}{r_{ij}}\right)^{n}\,.
  \label{eq:potential}
\end{equation}
This includes the standard 12-6 LJ potential, but also, e.g., potentials with more than two power-law terms. We note that some systems interacting via Eq.~(\ref{eq:potential}) will \emph{not} be strongly correlating and thus not have good isomorphs. In the following properties are derived for those systems that $do$ have good isomorphs. 


The total potential energy of a given micro-configuration $\mathbf R$ at density $\rho_*$ is a sum over contributions from the power-law terms, $U = \sum_n U_{n}$. When scaling  $\mathbf R$ to the density  $\rho$, keeping the structure invariant in reduced units, each power-law term is scaled by $\tilde\rho^\frac{n}{3} = (\rho/\rho_*)^{n/3}$, and the potential energy at the new density $U'=U\left( \tilde\rho^{-1/3}  \mathbf R \right)$ is:\cite{paperV}
 \begin{equation}
  U' = \sum_{n} \tilde\rho^\frac{n}{3}U_{n}.
\end{equation}
Thus the linear regression slope of the $U',U$-scatter plot in the direct isomorph check is given by (where all averages refer to the reference state point $(\rho_*,T_*)$):
\begin{equation}
\frac{\langle \Delta U' \Delta U  \rangle}{\langle (\Delta U)^2 \rangle} \, = 
   \sum_{n} \tilde\rho^{\frac{n}{3}}\frac{\langle \Delta U_{n} \Delta U  \rangle}{\langle (\Delta U)^2 \rangle}.
  \label{eq:grho}
\end{equation}
Using Einstein's fluctuation formula for the excess isochoric heat capacity and the corresponding formula for the ``partial'' heat capacities (which can be negative),
\begin{equation}
  C^{ex}_{v,n} \equiv \left(\frac{\partial \left< U_{n} \right>}{\partial T} \right)_V = \frac{\langle \Delta U_{n} \Delta U  \rangle}{k_BT^2}\,,
\end{equation}
we get an expression for the new temperature $T$ relative to the reference temperature $T_*$ (compare Eq.~(\ref{Eq:DirectIsomorph})):
\begin{equation}
 \frac{T}{T_*} = \frac{\langle \Delta U' \Delta U  \rangle}{\langle (\Delta U)^2 \rangle} \, =\sum_n \tilde\rho^{\frac{n}{3}}\frac{C^{ex}_{v,n}}{C^{ex}_v}\equiv  g(\tilde\rho)\,.
  \label{eq:grhoLJ}
\end{equation}
Since $C^{ex}_v = \sum_{n}C^{ex}_{v,n}$ the number of parameters in the scaling function $g(\tilde\rho)$ 
is one less than the number of terms in the potential (Eq.~(\ref{eq:potential})). 
In particular, for the standard 12-6 LJ potential,  $g(\tilde\rho)$ contains just a single parameter: 
\begin{equation}
 g(\tilde\rho) = \tilde\rho^4 c + \tilde\rho^2 \left(1 - c\right), ~~~ c \equiv {C^{ex}_{v,12}}/{C^{ex}_v}. 
 \label{eq:grho126LJ}
\end{equation}
Using that $U_{12} = W/2 - U$ for 12-6 LJ systems,\cite{paperV} $g(\tilde\rho)$  can conveniently be expressed in terms of $\gamma_*$ defined as Eq.~(\ref{eq:fluctgamma}) evaluated at the reference density $\rho_*$:
\begin{equation}
 g(\tilde\rho) = \tilde\rho^4 \left(\gamma_*/2 -1\right) - \tilde\rho^2 \left(\gamma_*/2 - 2 \right). 
 \label{eq:grho126LJ_gamma}
\end{equation}

$g(\tilde\rho)$ provides a new and convenient method for numerically tracing out an isomorph -- previously this could only be done by changing density by a small amount, e.g., 1\%, and then adjusting temperature to keep the excess entropy constant, using that $\gamma$ (Eq.~(\ref{eq:fluctgamma})) also can be expressed as:\cite{paperIV}  
\begin{equation}
 \gamma=\left(\frac{\partial\ln T}{\partial\ln\rho} \right)_{S_{ex}}. 
\label{eq:gamma_diffeq}
\end{equation}

It is a prediction of the isomorph theory that $\gamma$ depends only on density.\cite{paperIV,sch11} This means that the same differential equation, Eq.~(\ref{eq:gamma_diffeq}), determines the temperature on all isomorphs, implying that $g(\tilde\rho)$ is the same for all isomorphs -- what changes between different isomorphs is $T_*$. Thus $g(\tilde\rho)/T$ is an isomorph invariant (compare Eq.~(\ref{eq:grhoLJ})), which can be used as a scaling variable for the reduced relaxation time $\tilde\tau$ that is also an isomorph invariant\cite{paperIV} $\tilde\tau =f(g(\tilde\rho)/T)$. This form of scaling  was first proposed by Alba-Simionesco {\it et al.}\cite{alb} Here a theoretical derivation has been provided, as well as an explicit expression for $g(\tilde\rho)$ for systems interacting via generalized LJ potentials (Eq.~(\ref{eq:potential})). We note further that since the solid-liquid coexistence lines of strongly correlating liquids are predicted to be isomorphs\cite{paperIV,paperV}, this immediately explains the recent observation of Khrapak and coworkers that the solid-liquid coexistence line of an LJ liquid is given by $\left( C_1 \rho^4 - C_2 \rho^2\right)/T={\rm Const.}$\cite{kra}


From the scaling $\tilde\tau=f(g(\tilde\rho)/T)$ follows that \emph{pairs} of isochores obey power-law density scaling, since a power-law $A\rho^\gamma$ that is equal to $g(\tilde\rho)$ at the two densities involved always exists. This is indeed what is seen in Fig.~\ref{fig:ds_break}.
Choosing $\rho_*=1.6$ as reference density, the scaling in the left panel of Fig.~\ref{fig:ds_break} corresponds to $g(1.2/1.6) = (1.2/1.6)^{4.90}$ which via Eq.~(\ref{eq:grho126LJ_gamma}) leads to $\gamma_*=4.59$. Using this value we find $g(2.0/1.6) = 2.70 = (2.0/1.6)^{4.45}$. Thus from one scaling exponent in Fig.~\ref{fig:ds_break}, the other is uniquely predicted. Moreover, the value $\gamma_*=4.59$ is consistent with what is found by evaluating Eq.~(\ref{eq:fluctgamma}) at the reference isochore $\rho_*=1.6$ in the temperature range $T=1.7$ to $5$, which leads to values of $\gamma_*$ decreasing from 4.6 to 4.5. In the following figures reporting results for the KABLJ mixture, we use $\gamma_*=4.59$ as estimated from the left panel of  Fig.~\ref{fig:ds_break}, i.e., no further fitting or adjustment of parameters was applied.

\begin{figure}
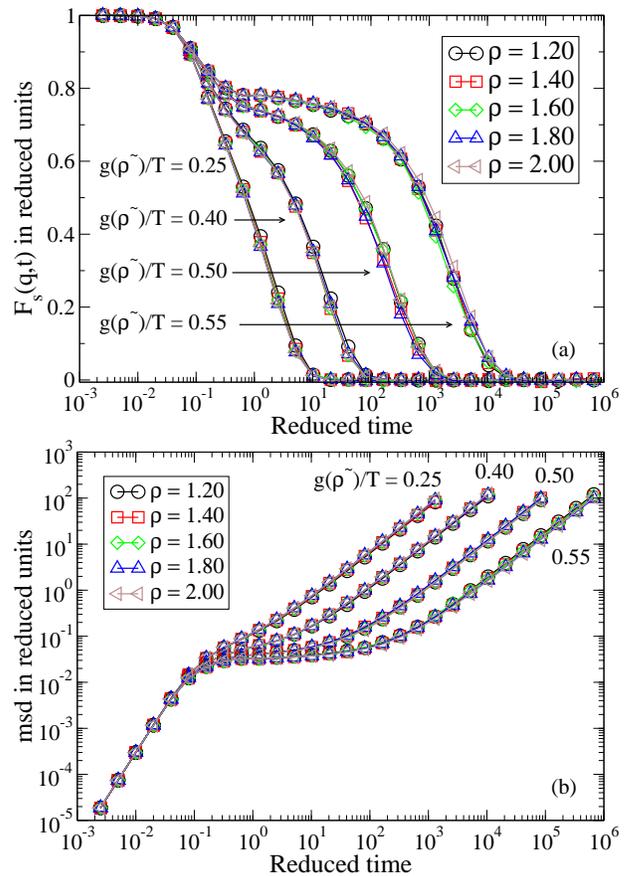

   \includegraphics[width=0.45\textwidth]{fig2a_v2.eps}
   \includegraphics[width=0.45\textwidth]{fig2b_v2.eps}
   \caption{Four different isomorphs in the KABLJ mixture, each generated from the condition $g(\tilde\rho)/T={\rm Const.}$ Densities range from 1.2 to 2.0, and $g(\tilde\rho)=\tilde\rho^4 \left(\gamma_*/2 -1\right) - \tilde\rho^2 \left(\gamma_*/2 - 2 \right)$  (Eq.~(\ref{eq:grho126LJ_gamma})) with $\gamma_*=4.59$ determined from the scaling in Fig.~\ref{fig:ds_break}, see text ($\tilde{\rho} \equiv \rho/\rho_*$, $\rho_*=1.6$). (a) Self part of intermediate scattering functions in reduced units. (b) Mean-square displacements in reduced units. The data collapse confirms that true isomorphs have been identified.
}
    \label{IM_g}
\end{figure}

As mentioned, the scaling function $g(\tilde\rho)$ was derived \emph{assuming} that good isomorphs exist. In Fig. \ref{IM_g} we test this for the KABLJ mixture using the most sensitive isomorph invariant - the dynamics of viscous states.  State points with the same $g(\tilde\rho)/T$, predicted to be on the same isomorph, are seen to have very similar dynamics even though density varies from 1.2 to 2.0.

\begin{figure}
   \includegraphics[width=0.45\textwidth]{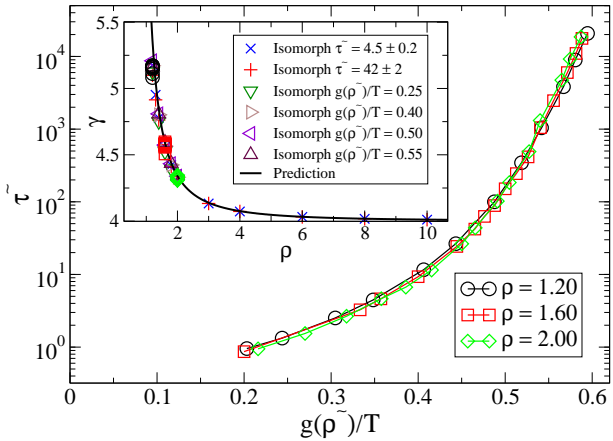} 
   \caption{Reduced relaxation times for the KABLJ mixture scaled according to the isomorph theory (same data as in Fig.~\ref{fig:ds_break}). The scaling function $g(\tilde\rho)$ is the same as in Fig.~\ref{IM_g} (Eq.~(\ref{eq:grho126LJ_gamma}), $\gamma_*=4.59$).  
Inset: Comparing $\gamma$ computed from simulations (Eq.~(\ref{eq:fluctgamma})) to the prediction of the isomorph theory, $\gamma=d\ln g /d\ln\rho$ (black curve). Isomorphs denoted by a reduced relaxation time are each generated from a single reference point using Eqs.~(\ref{eq:grhoLJ}) and~(\ref{eq:grho126LJ}), with '$\pm$' quantifying the resulting variation of the  reduced relaxation time on the isomorph.
Isochores are plotted with the same symbols as in the main figure. 
}
   \label{g_scal}
\end{figure}
Figure \ref{g_scal} tests the proposed scaling for the KABLJ mixture using the data of Fig.~\ref{fig:ds_break}. Clearly the new form of scaling works well. Combining Eq.~(\ref{eq:gamma_diffeq}) with the definition of $g(\tilde\rho)$ (Eq.~(\ref {eq:grhoLJ})) shows that $\gamma$ is the logarithmic derivative of $g(\tilde\rho)$, $\gamma=d\ln g /d\ln\rho$. The inset of Fig.~\ref{g_scal} demonstrates that this prediction agrees well with simulations, even when going to large densities where the purely repulsive $r^{-12}$  limit is approached. 


\begin{figure}
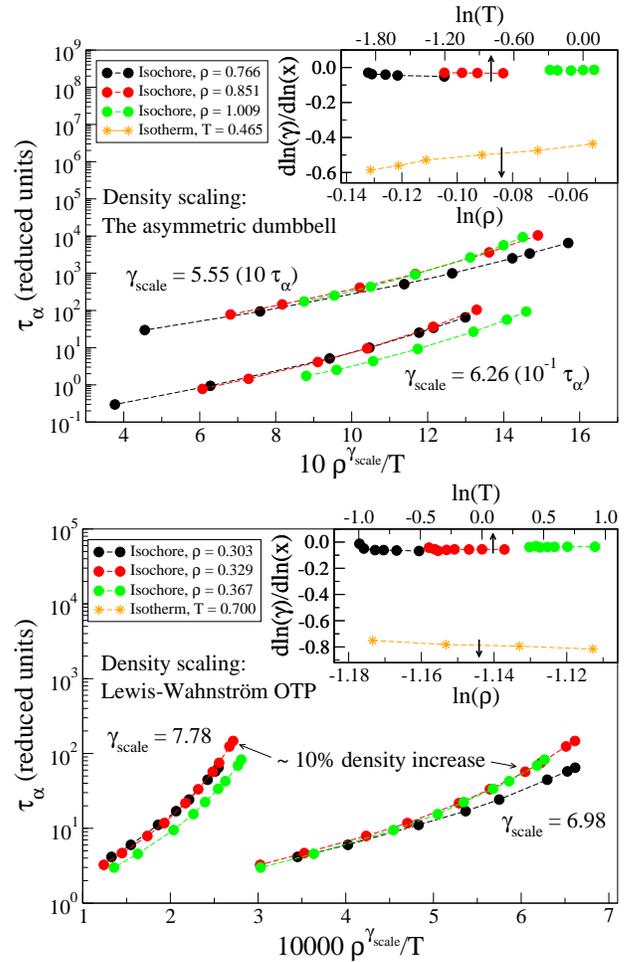

  \includegraphics[clip=true,width=0.45\textwidth]{fig4a.eps}
  \includegraphics[clip=true,width=0.45\textwidth]{fig4b.eps}
   \caption{Breakdown of power-law density scaling in two molecular models. In accordance with the scaling derived in the present work, power-law scaling \emph{does} work when considering only pairs of isochores. a) Asymmetric dumbell model. b) Lewis-Wahnstrom model of ortho-terphenyl (OTP). Insets:  $(\partial\ln \gamma /\partial\ln T)_{\rho}$ plotted against $\ln T$ (circles) and $(\partial\ln \gamma /\partial\ln \rho)_{T}$ plotted against $\ln \rho$ (stars). Consistent with the isomorph theory, $\gamma$ is found to be much more dependent on density than on temperature.
    }
   \label{molecules}
\end{figure}

We now turn briefly to model molecular systems. In this case it is still a prediction of the isomorph theory that an expression of the form $g(\tilde\rho)/T$ is the right scaling variable \cite{sch11}, but we do not have an explicit expression for $g(\tilde\rho)$. Figure \ref{molecules} demonstrates how power-law density scaling breaks down for the Lewis-Wahnstrom model of ortho-terphenyl and an asymmetric dumbell model when considering larger density changes than previously studied.\cite{sch09} Like in Fig.~\ref{fig:ds_break}, power-law scaling works when considering pairs of isochores, consistent with the right scaling variable being of the form $g(\tilde\rho)/T$. The insets of Fig.~\ref{molecules} test the isomorph prediction that $\gamma$ to a good approximation is a function of density only, the assumption used to derive the new scaling. The prediction agrees well with simulations: $\gamma$ is found to be much more dependent on density than on temperature. For more results on isomorphs in these model molecular liquids see Ref.~\onlinecite{ing11}.

\begin{figure*}
    \includegraphics[width=0.42\textwidth]{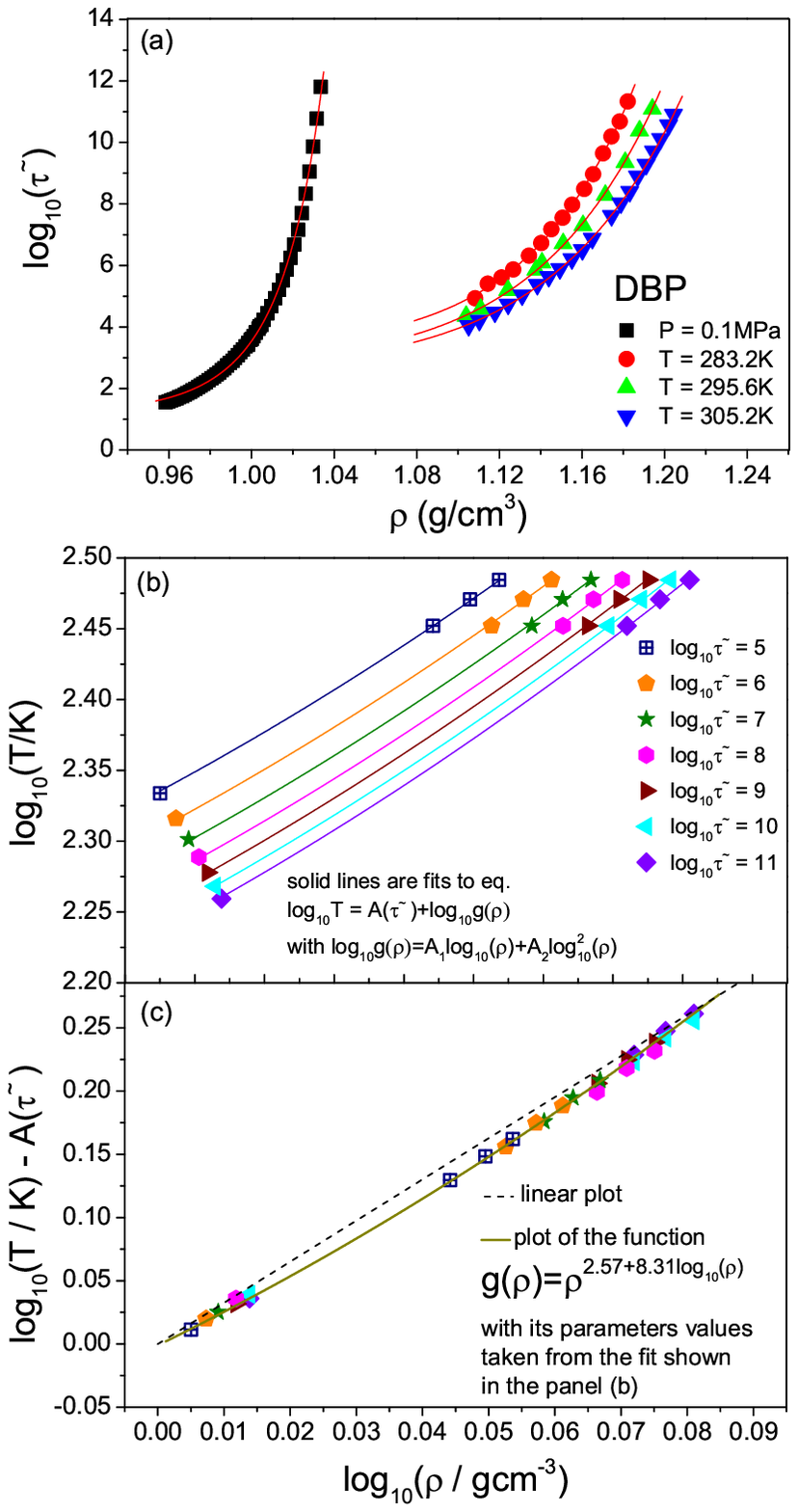} 
    \includegraphics[width=0.42\textwidth]{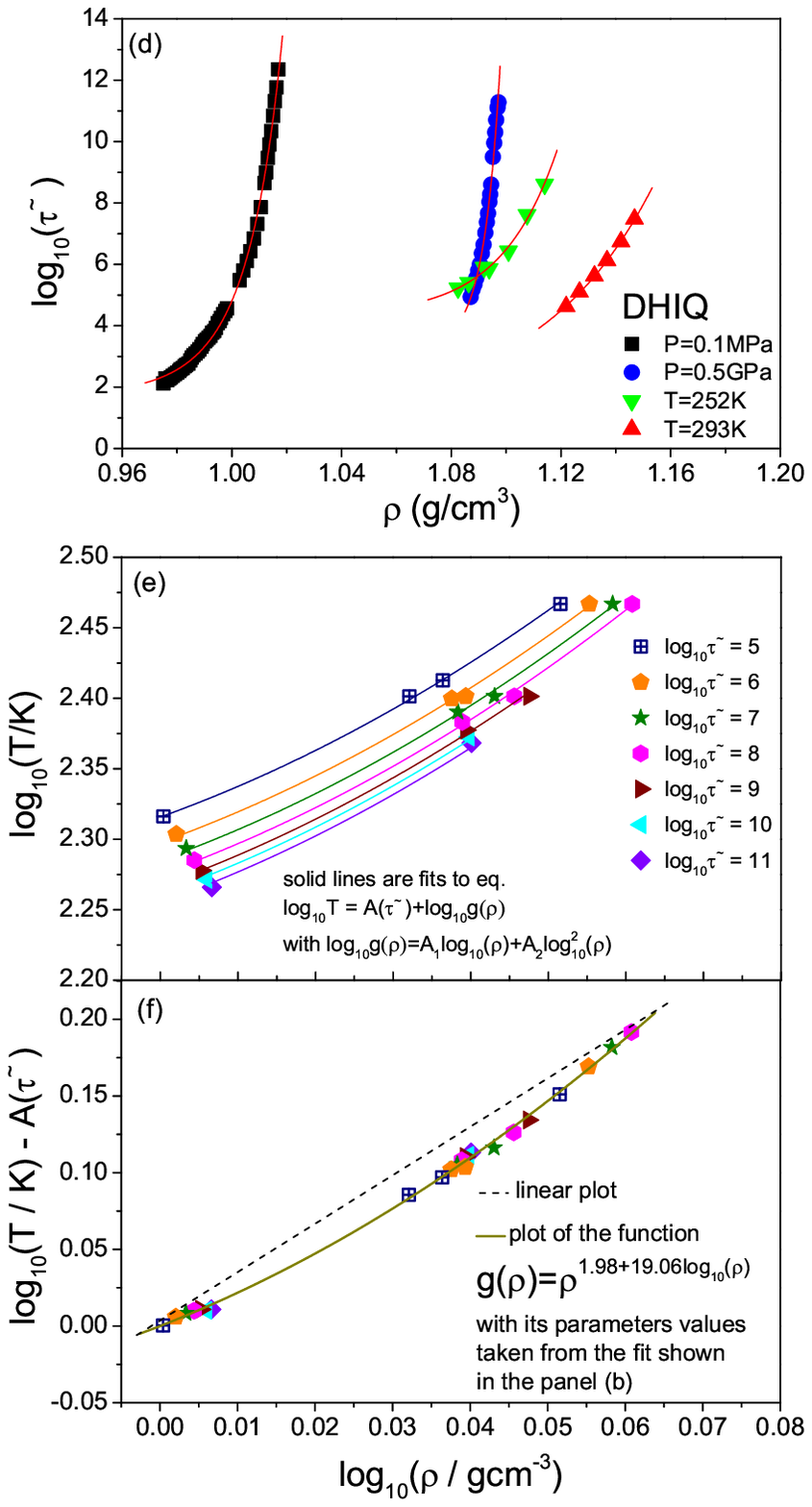} 
    \caption{Deviation from  power-law scaling in DBP (a-c) and DHIQ (d-f). (a) and (d) Density dependence of isobaric and isothermal structural relaxation times $\tau\tilde{}$ in reduced units. Solid lines represents separate fits to the modified Avramov model.\cite{cas06b} (b) and (e) The isochronal dependences $\log_{10}T$ vs $\log_{10}\rho $ determined at given $\tau\tilde{}$. Fits are done to all isochrones simultaneously.
 (c) and (f) The isochrones vertically shifted by the fitted values $A(\tau \tilde{})$. The deviations from power-law scaling (straight dashed lines) are evident. The fitted $g(\rho)$'s corresponds to $\gamma$ values increasing from 2.6 to 3.9 (DBP) and from 2.0 to 4.3 (DHIQ).
 }
    \label{fig:DBP_DHIQ}
 \end{figure*}

Finally, we present experimental results for the two van der Waals liquids dibutyl phthalate (DBP) and decahydroisoquinoline (DHIQ), using larger  density increases than usually studied in scaling experiments (20\% and 18\% respectively). Using earlier reported dielectric data for DBP\cite{DBP} and  DHIQ\cite{pal05} and values of the Tait equation of state parameters for DBP\cite{rol06} and  DHIQ\cite{cas06}, we find that the isochronal dependences $\log_{10}T$ vs $\log_{10}\rho $ determined at given structural relaxation times in reduced units, $\tau\tilde{}  = \tau v_{m}^{-1/3}(k_{B}T/m)^{1/2}$, where $v_{m}$ and $m$ are molecular volume and mass, are nonlinear (Fig.~\ref{fig:DBP_DHIQ}(b) and (e)). This implies breakdown of power-law density scaling. The isochrones can be superposed after vertical shifting, however (Fig.~\ref{fig:DBP_DHIQ}(c) and (f)), which implies that a scaling variable exists of the form $g(\rho)/T$. The isochrones can be described by a phenomenological form of the scaling function $\log_{10}g(\rho ) = A_{1}\log_{10}\rho + A_{2}\log_{10}^2\rho$, chosen simply to take into account the curvature. 
In the case of DBP, our results are consistent with those reported by Niss et al.\cite{niss07}
We conclude that for DBP and DHIQ power-law density scaling breaks down at large density variations in the way predicted by the isomorph theory. Interestingly, the density dependence of $\gamma$ is stronger than for the model systems, and in the opposite direction; for the experimental systems $\gamma$
 \emph{increases} with density. 


What are the perspectives of our findings? Based on the theory of isomorphs in dense liquids we have now  a form of density scaling that is more fundamental and more robust than power-law density scaling, and which is consistent with both simulations and experiments. This ``isomorph scaling'' -- originally proposed by Alba-Simionesco {\it et al.}\cite{alb} -- is predicted to apply for all strongly correlating liquids, i.e., van der Waals and metallic liquids, but not, e.g., for hydrogen-bonding liquids. Our results should \emph{not} be used to abandon power-law density scaling  -- it is a useful approximation to isomorph scaling when the scaling function $g(\rho)$ is unkown and only moderate density changes are considered. Under these conditions the isomorph theory predicts that power-law density scaling works with an exponent determined by Eq.~(\ref{eq:fluctgamma}). Isomorph scaling provides a deeper understanding of why -- and when -- power-law density scaling works.

Isomorph scaling has important consequenses regarding the most fundamental open question in the field of  viscous liquids and the glass transition: what controls the dramatic viscous slowing down? In general this question has to be considered as a search for a physically justified function of {\it two} variables, temperature and density (or temperature and pressure). Our results imply that this problem is now effectively reduced to a search for a function of a single variable, at least for the class of strongly correlating liquids. This is particularly striking for LJ type systems like the KABLJ mixture, where we have a prediction for the scaling function, that  agrees very well with simulations for much larger density variations than usually considered. The fact that the LJ scaling function contains just a single parameter -- i.e., no more parameters than power-law density scaling -- confirms that isomorph scaling is more fundamental and not merely a higher-order approximation compared to  power-law density scaling. Isomorph scaling must be taken into account in theories of the viscous slowing down: since the relaxation time in reduced units obeys isomorph scaling, any quantity proposed to control the relaxation time must also obey isomorph scaling.


\acknowledgments 
The centre for viscous liquid dynamics ``Glass and Time'' is sponsored by the Danish National Research Foundation (DNRF). A. G. and M. P. gratefully acknowledge the support from the Polish National Science Centre   (project no. N N202 023440).


\end{document}